\documentclass[prb,groupedaddress,twocolumn]{revtex4-2}

\usepackage{graphicx}
\usepackage{dcolumn}
\usepackage{bm}
\usepackage{hyperref}
\usepackage{url}
\usepackage{color}
\usepackage{amsmath}
\usepackage[running]{lineno}

\newcommand{\pcto}{PbCuTe$_{2}$O$_{6}$\,}
\newcommand{\scto}{SrCuTe$_{2}$O$_{6}$\,}

\begin{document}
\title{Magnetic structure of a new quantum magnet \scto}

\author{S. Chillal}
\email[*]{shravani.chillal@helmholtz-berlin.de}
\affiliation{Helmholtz-Zentrum Berlin f\"ur Materialien und Energie, Hahn-Meitner Platz 1, 14109 Berlin, Germany}

\author{A. T. M. N. Islam}
\affiliation{Helmholtz-Zentrum Berlin f\"ur Materialien und Energie, Hahn-Meitner Platz 1, 14109 Berlin, Germany}

\author{H. Luetkens}
\affiliation{Laboratory for Muon Spin Spectroscopy, Paul Scherrer Institut (PSI), 5232 Villigen, Switzerland}

\author{E. Can\'evet}
\affiliation{Laboratory for Neutron Scattering and Imaging, Paul Scherrer Institut (PSI), 5232 Villigen, Switzerland}
\affiliation{Department of Physics, Technical University of Denmark, 2800 Kongens Lyngby, Denmark}

\author{Y. Skourski}
\affiliation{Dresden High Magnetic Field Laboratory (HLD-EMFL), Helmholtz-Zentrum Dresden-Rossendorf, D-01328 Dresden, Germany}

\author{D. Khalyavin}
\affiliation{ISIS Facility, STFC Rutherford Appleton Laboratory, Oxfordshire OX11 0QX, UK}

\author{B. Lake}
\affiliation{Helmholtz-Zentrum Berlin f\"ur Materialien und Energie, Hahn-Meitner Platz 1, 14109 Berlin, Germany}
\affiliation{Institut f\"ur Festk\"orperphysik, Technische Universit\"at Berlin, Hardenbergstr. 36, 10623 Berlin, Germany}

\begin{abstract}
\scto consists of a 3-dimensional arrangement of spin-$\frac{1}{2}$ Cu$^{2+}$ ions. The 1$^{st}$, 2$^{nd}$ and 3$^{rd}$ neighbor interactions respectively couple Cu$^{2+}$ moments into a network of isolated triangles, a highly frustrated hyperkagome lattice consisting of corner sharing triangles and antiferromagnetic chains. Of these, the chain interaction dominates in \scto while the other two interactions lead to frustrated inter-chain coupling giving rise to long range magnetic order at suppressed temperatures. In this paper, we investigate the magnetic properties in \scto using muon relaxation spectroscopy and neutron diffraction and present the low temperature magnetic structure.   
\end{abstract}


\date{\today}

\maketitle

Interesting magnetic behaviour in Heisenberg spin systems originates from a network of some elementary motifs such as triangles or tetrahedra, where spins at their vertices interact with each other via antiferromagnetic (AF) interactions. The frustration in such systems often leads to exotic ground states such as spin liquids~\cite{Han2012,Balz2016} and spin ice states~\cite{Benton2012,Gingras2014} where long-range magnetic order (LRO) is suppressed to low temperatures or completely eliminated. In the case where order still occurs it can provide insights into the underlying physics and the new states arising from the frustration. There are many experimental examples for the three dimensional (3D) networks of corner-shared tetrahedra (pyrochlore~\cite{Benton2012,Gingras2014,Clark2014} and spinel structures~\cite{Higo2017,Chamorro2018}) such as Gd$_{2}$Hf$_{2}$O$_{7}$~\cite{Durand2008}, 3D networks of corner-shared triangles are relatively less explored despite the expectation of novel ground states. The simplest possibility of the latter is known as a hyperkagome lattice and has been observed in the compound Na$_{4}$Ir$_{3}$O$_{8}$ where every Ir$^{2+}$ spin is involved in two triangles. Although initial studies suggested a highly frustrating magnetic lattice with QSL behaviour~\cite{Zhou2008}, a glassy magnetic ground state has been observed in the muon relaxation studies~\cite{Bergholtz2010,Dally2014}.

\pcto is an example of a highly connected hyperkagome lattice, also known as the hyper-hyperkagome lattice, formed by the highly frustrated first and second nearest neighbour (NN) interactions between Cu$^{2+}$ spins~\cite{Chillal2020}. Experimental and theoretical studies of this compound reveal evidence for quantum spin liquid behaviour down to 20~mK, a rare observation in three dimensional magnetic lattices~\cite{Koteswararao2014,Khuntia2016,Chillal2020}, confirming the strong frustration in the system. However, density functional theory calculations also suggest significant non-frustrated third and fourth NN magnetic interactions in \pcto whose role in the QSL phase diagram is less understood. 

\scto is a promising quantum magnet, iso-structural to \pcto, that can give insights into the hyper-hyperkagome frustration mechanism responsible for the QSL ground state. \scto crystallizes in cubic symmetry at room temperature (space group {\it P4$_{1}$32}~\cite{Wulff1997}) with the magnetic spin-$\frac{1}{2}$ Cu$^{2+}$ ions occupying a single Wyckoff site. The Cu$^{2+}$ ions are coupled together by exchange interactions $J_{1}$, $J_{2}$ and $J_{3}$. These three interactions couple them into isolated equilateral triangles, a hyperkagome lattice and uniform chains (running parallel to the {\it a}, {\it b} and {\it c} axes) respectively. If these interactions are antiferromagnetic they can give rise to a frustrated network of spin-$\frac{1}{2}$ chains. DC susceptibility of \scto yields a negative Curie-Weiss temperature of $\theta_{CW}\approx-35.4$~K revealing predominantly antiferromagnetic exchange interactions~\cite{Koteswararao2015,Ahmed2015}, and shows a broad maximum at 32~K. This feature has been attributed to a one-dimensional spin-$\frac{1}{2}$ Heisenberg antiferromagnetic chain revealing $J_{3}=-45$~K~\cite{Koteswararao2015,Ahmed2015} as the dominant interaction. However, two sharp features occur in the susceptibility at lower temperatures  $T_{N1}=5.5$~K and $T_{N2}=4.5$~K, where a sharp $\lambda$-type anomaly is also observed in the heat capacity, indicating the onset of magnetic transitions in the system. These anomalies reveal non-negligible frustrated inter-chain coupling due to the finite $J_{1}$ and $J_{2}$~\cite{Koteswararao2015,Ahmed2015}. In addition, the compound exhibits magneto-dielectric coupling at $T_{N1}$ and $T_{N2}$~\cite{Koteswararao2016} attributed to the non-centro-symmetric nature of the structural symmetry. Furthermore, specific heat, magnetization and dielectric constant measurements as a function of applied magnetic field reveal a complex phase diagram with an additional field induced phase~\cite{Koteswararao2015,Ahmed2015}. 

Although \scto reveals interesting magneto-dielectric and magnetoelectric properties around the magnetic transitions, the origins of the magnetic order and the nature of the magnetic structure below the transition temperatures is not known. Here, we present the field-temperature phase diagram for three different directions of the single crystalline samples of \scto that shed light on the magnetic properties of the compound. Further, we investigate the polycrystalline samples with muon spin resonance ($\mu^{+}$SR) and neutron powder diffraction measurements and propose a model for the zero-field magnetic structure in the ordered state. The results reveal that the first neighbor triangle interaction provides the interchain coupling and is responsible for the long-range order in the system.

\section{Samples \& Experimental Methods}

\begin{figure}
\includegraphics[width=1 \columnwidth]{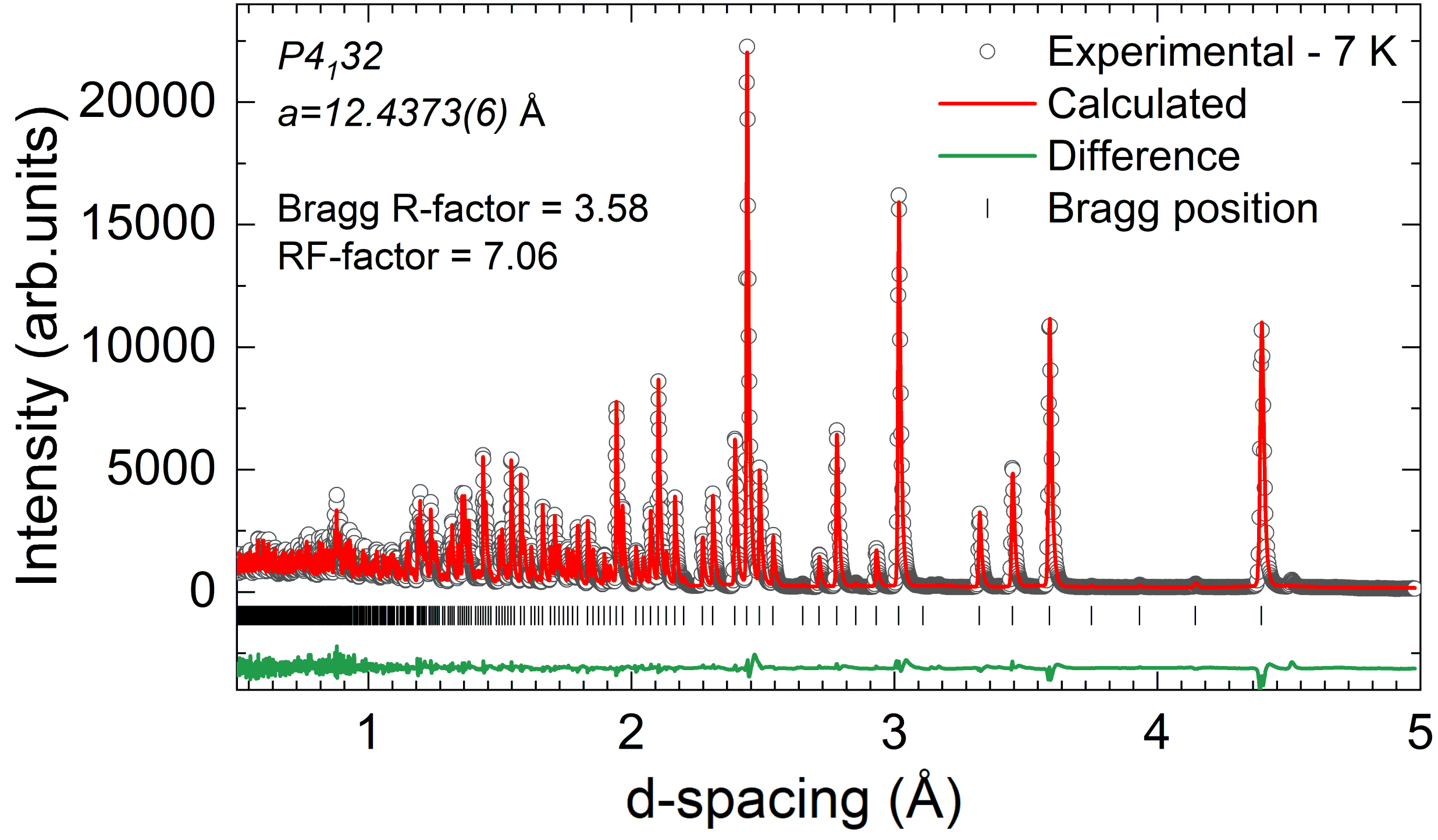}
\caption{Neutron powder diffraction pattern of \scto measured in the paramagnetic state at $T=7$~K on the WISH diffractometer at a mean 2$\theta=154^{\circ}$. The pattern can be well fitted by considering a cubic structure ( {\it P4$_{1}$32} space group) and lattice constant of 12.4373~\AA{} using Rietveld refinement.}
\label{Fig1}
\end{figure}

\begin{table}
\centering
\begin{tabular}{|c|c|c|c|c|c|} 
\hline
\hline
\bf Atom & \bf \begin{tabular}[c]{@{}l@{}}Wyckoff\\position\end{tabular} & \bf x/a & \bf y/a & \bf z/a & \bf B$_{iso}$ \\ 
\hline
Te          & 24e & 0.33775  & 0.91970 & 0.05890 & 0.46001   \\ 
\hline
Sr1         & 8c   & 0.05335 & 0.05335 & 0.05335 & 0.65537   \\ 
\hline
Sr2         & 4b   & 0.87500 & 0.87500 & 0.87500 & 0.61456   \\
\hline
Cu          &12d  & 0.12500 & 0.77446 & 0.02445 & 0.47196   \\
\hline
O1         & 24e  & 0.57936 & 0.92944 & 0.37654 & 0.25773   \\
\hline
O2         & 24e  & 0.26670 & 0.81156 & 0.97806 & 0.49215   \\
\hline
O3         & 24e  & 0.22239 & 0.97760 & 0.12925 & 0.53796   \\
\hline
\end{tabular}
\caption{The Rietveld refined coordinates and isotropic thermal parameters of \scto at 7~K.}
\label{tab0}
\end{table}

Polycrystalline powder of \scto was prepared from stoichiometric mixture of high purity powders of SrCO$_{3}$ (99.99\%), CuO (99.995\%) and TeO$_{2}$ (99.99\%) by solid state reactions at 650$^\circ$C in a vacuum furnace under Argon flow.
For crystal growth, first stoichiometric amounts of high purity SrCO$_{3}$, CuO and TeO$_{2}$ were mixed as above and  sintered twice for 12 hours at 600$^\circ$C in Argon flow with intermediate grinding. Then a feed rod (diameter$\approx$6~mm, length$\approx$7-8~cm) was prepared from the stoichiometric powder and densified by pressing in a Cold Isostatic press in 2000~bars and subsequent sintering at 650$^\circ$C in Argon flow. Crystal growth was done using the feed-rod by the Floating zone technique in a four mirror type optical image furnace (Crystal Systems Corp., Japan). Growth was done at a rate of 1~mm/hr in Argon atmosphere at ambient pressure. The as-grown crystal is approximately 5~mm diameter and 3.5~cm in length. It was checked by X-ray Laue diffraction for single crystallinity and confirmed by polarized optical microscopy to be free of inclusions. The quality of the crystal has also been analyzed for phase purity by grinding a small piece of the crystal into powder upon which x-ray diffraction was performed. These single crystals reveal a small quantity of non-magnetic impurity in the form of Sr$_{2}$Te$_{3}$O$_{8}$ amounting to less than 1\%. The single crystals were then characterized by magnetic susceptibility, magnetization and heat capacity in the temperature range of 1.8$-$400~K and an external field of 0$-$7~T using a Physical Property Measurement System (PPMS). The sample synthesis and characterization took place at the Core Lab for quantum Materials, Helmholz-Zentrum Berlin, Germany.

 $\mu^{+}$SR measurements on the polycrystalline \scto were performed at the General purpose Spectrometer (GPS) at the SMuS facility in Paul Scherrer Institut down to 1.6~K in zero field. The nuclear and magnetic structure of \scto was investigated between 20~K and 1.6~K by obtaining neutron diffraction patterns on powder sample of 10~g. An initial search for the magnetic Bragg peaks was carried out at the DMC diffractometer~\cite{Scheffer1990} at the Paul Scherrer Institut, Switzerland using two incident wavelengths $\lambda=2.46$~\AA{} and 4.504~\AA{} (PG002 monochromator) covering a momentum transfer Q in the range of $0.2$~\AA$^{-1}<$Q$<~3.7$~\AA$^{-1}$ and $0.35$~\AA$^{-1}<~$Q$~<2$~\AA$^{-1}$ respectively. The diffraction patterns were collected at 1.6~K, 5.2~K and 20~K. Detailed temperature dependence of the nuclear and magnetic structure on the powder sample was performed at the time-of-flight diffractometer WISH~\cite{Chapon2011} at the ISIS facility, UK. The patterns were collected for temperatures between 1.5~K and 15~K and momentum transfer $0.37$~\AA$^{-1}<$Q$<9$~\AA$^{-1}$. In both cases, the powder was loaded into a cylindrical vanadium can and the temperature was controlled using a typical orange cryostat. The patterns are refined using the Rietveld method in the Fullprof package~\cite{Rodriguez1993} and magnetic symmetry analysis was performed using a combination of BasiReps and Bilbao crystal server software packages~\cite{Aroyo2011}. Figure.~\ref{Fig1} shows the neutron powder diffraction of the nuclear structure taken at 7~K at the WISH diffractometer. The refinement agrees with the non centro-symmetric cubic structure {\it space group: P4$_{1}$32}, consistent with previously reported results~\cite{Koteswararao2015,Ahmed2015} at room temperature. The lattice constant at 7~K is found to be 12.4373(2)~\AA. The refined values of the coordinates and thermal factors are listed in Table.~\ref{tab0}. 

\section{Results}

\subsection{Magnetic properties of single crystal}

\begin{figure}
\includegraphics[width=1 \columnwidth]{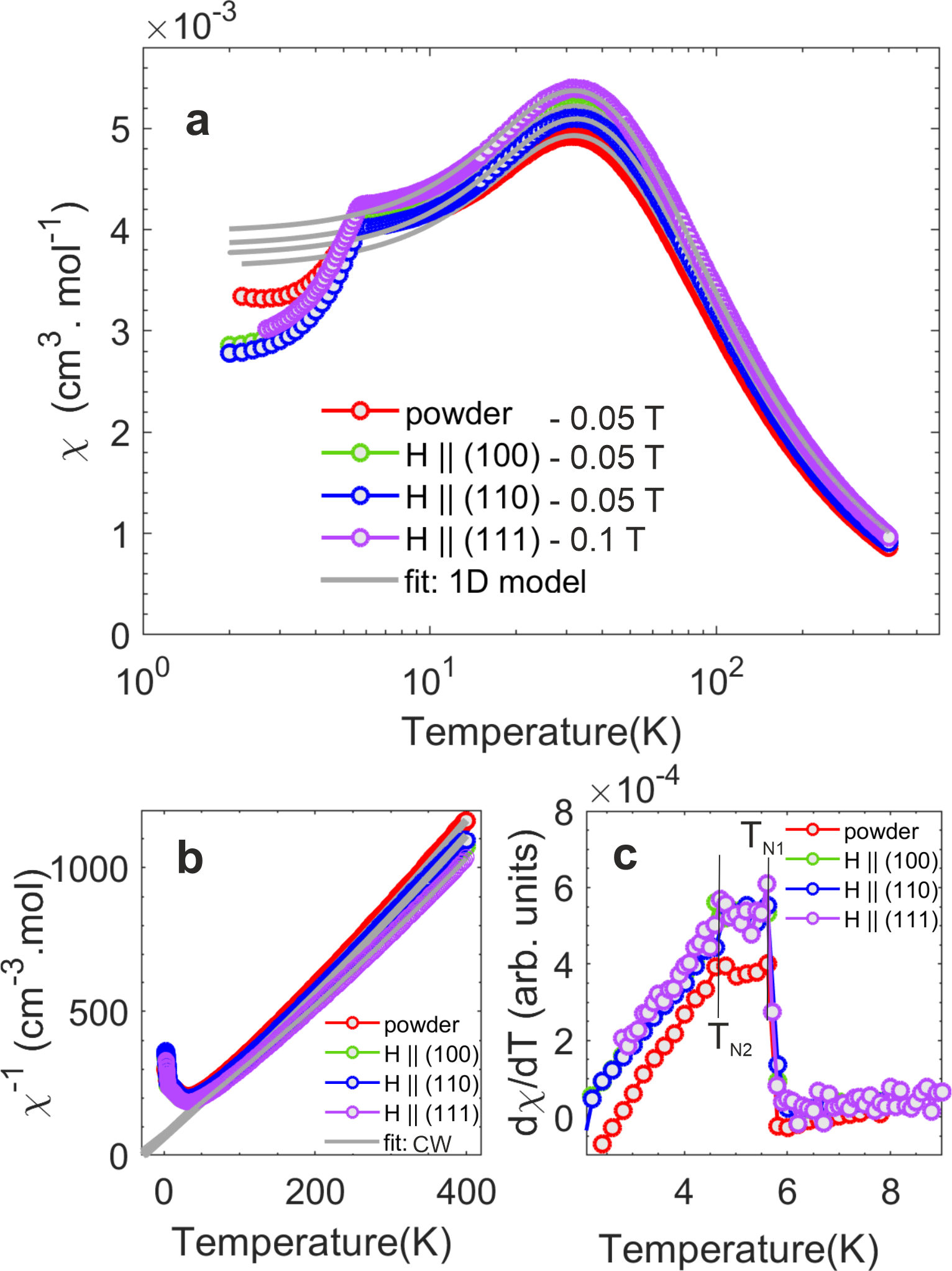}
\caption{\textbf{a}) Susceptibility of polycrystalline and single-crystal samples of \scto exhibiting a broad hump at $\sim32$~K. The solid lines are fits to the numerical antiferromagnetic spin-$\frac{1}{2}$ chain susceptibility~\cite{Eggert1994,Johnston2000}, \textbf{b}) Curie-Weiss fit to the inverse of the susceptibility. \textbf{c}) Derivative of dc-susceptibility (shown in panel \textbf{a}) for the single crystal and polycrystalline samples revealing two anomalies at $T_{N1}\approx5.5$~K and $T_{N2}\approx4.5$~K.}
\label{Fig2a}
\end{figure}

Figure.~\ref{Fig2a}\textbf{a} shows the zero-field-cooled dc-magnetic susceptibility of the polycrystalline and single crystal samples in a bias field of H$=0.05$~T revealing several important clues to the magnetic state of the system (1.8~K-400~K). At high temperatures, the inverse susceptibility is linear (fig.~\ref{Fig2a}\textbf{b}) and can be fitted to paramagnetic Curie-Weiss (CW) behaviour: $\chi=\chi_{core}+\chi_{vv}+\frac{C}{T-\theta_{CW}}$, where $\chi_{core}=-1.54\times10^{-4}$~cm$^{3}$. mol$^{-1}$ is the diamagnetic contribution from the core non-magnetic ions Te$^{4+}$ ions and $\chi_{vv}$ refers to Van Vleck paramgnetism. In order to obtain reliable values of the Curie-Weiss temperature $\theta_{CW}$, we have varied the lower bound of the temperature range of the fits from 100~K to 200~K. The best fits are obtained for 140~K$-$400~K and the resulting fit parameters $\chi_{vv}$, Curie-Weiss constant $C$, $\theta_{CW}$ along with the derived $\mu_{eff}=3Ck_{B}N_{A}/\mu_{B}$ and $g-$factor are tabulated in Table.~\ref{tab1}. The values of $\theta_{CW}$ are: $-$28 $\pm$ 0.3~K, $-$28 $\pm$ 1~K, $-$26 $\pm$ 1~K and $-$27.5 $\pm$ 1.5~K for polycrystalline and crystalline (100), (110) and (111) axes respectively. Within the sensitivity of the measurement and of demagnetization effects due to the shape of the crystal, the single crystal susceptibility in all crystalline directions follows that of the polycrystalline sample hence confirming the isotropic nature of the Cu$^{2+}$ spins in \scto. Furthermore, the negative $\theta_{CW}$ values confirm the predominant antiferromagnetic interactions in the system. The effective moment calculated from the Curie-Weiss constant is $\sim$1.85$~\mu_{B}$ which is very close to the full moment of the free Cu$^{2+}$ spin. Accordingly, the derived $g-$factor is close to 2.1 in the four measurements assuming spin-$1/2$. We find that the $\theta_{CW}$ values are smaller than the previously reported $\theta_{CW}=-35$~K in polycrystalline samples~\cite{Koteswararao2015,Ahmed2015}. The discrepancy could be attributed to the sensitivity of the $\theta_{CW}$ to the fitted temperature range. 

\begin{table}
\centering
\begin{tabular}{|c|c|c|c|c|c|} 
\hline
\bf Sample                      & \bf \begin{tabular}[c]{@{}l@{}}$\bm{\chi}_{vv}$($\times10^{-5})$\\(cm$^{3}$/mol)\end{tabular} & \bf \begin{tabular}[c]{@{}l@{}}C\\(cm$^{3}\cdot$K/mol)\end{tabular} & \bf \begin{tabular}[c]{@{}l@{}}$\bm{\theta}_{CW}$\\(K)\end{tabular} & \bf \begin{tabular}[c]{@{}l@{}}$\bm{\mu}_{eff}$\\($\bm{\mu}_{B}$)\end{tabular} & \bf \begin{tabular}[c]{@{}l@{}}g-\\factor\end{tabular}\\ 
\hline
Powder   & 4.49 $\pm$ 0.01& 0.413 $\pm$ 0.008 & 28.44 $\pm$ 0.3 & 1.82 & 2.1   \\ 
\hline
(100)    & 6.95 $\pm$ 0.05& 0.436 $\pm$ 0.003 & 27.94 $\pm$ 1 & 1.87 & 2.16   \\ 
\hline
(110)    & 5.38 $\pm$ 0.06 & 0.426 $\pm$ 0.003 & 26.15 $\pm$ 1 & 1.85 & 2.13   \\
\hline
(111)    & 11.72 $\pm$ 1.1 & 0.421 $\pm$ 0.005 & 27.5 $\pm$ 1.5 & 1.84 & 2.12   \\
\hline
\end{tabular}
\caption{The Curie-Weiss temperature, effective moment, and the g-factor as derived from the Curie-Weiss fit to the high temperature magnetic susceptibility (T$>140$~K, H$=0.05$~T) of the powder sample and single crystal sample aligned parallel to external field along the (100), (110) and (111) directions. Note: The higher $\chi_{vv}$ along (111) is likely due to the paramagnetic background from teflon wrapped on the sample (not used for the directions).}
  \label{tab1}
\end{table}

In the intermediate temperature range, all the four data sets exhibit a broad hump around $\sim$32~K indicative of short-range magnetic correlations, characteristic of 1D Heisenberg spin-$\frac{1}{2}$ chain compounds. The solid grey lines in fig.~\ref{Fig2a}a are a fit (T$>$15~K) to the high-temperature series expansion for the DC susceptibility of a spin-$\frac{1}{2}$ Heisenberg antiferromagnetic chain~\cite{Eggert1994,Johnston2000}:

\begin{equation}
\begin{split}
&\chi=\chi_{core}+\chi_{vv}+\frac{N_{A}{~\mu_{B}}^{2}g^{2}}{4k_{B}T}\times\\
&\frac{1+0.08516x+0.23351x^{2}}{1+0.73382x+0.13696x^{2}+0.53568x^{3}}
\end{split}
\label{eq1}
\end{equation}

where J$_{chain}$ in $x={J_{chain}}/{T}$ is the chain interaction which is also the 3$^{rd}$ nearest-neighbour interaction in the case of \scto. The g-factor and $\chi_{vv}$ are also fitted within this model and the resulting parameters are tabulated in Table.~\ref{tab2}. The model yields a chain interaction J$_{chain}\sim49$~K and a g-factor of $\sim$2.2 in the single crystal. The observed g-factor, although slightly higher than the fully isotropic spin system, it is consistent with the values obtained from high temperature Curie-Weiss behaviour. In Heisenberg systems the Curie-Weiss temperature is the weighted sum of all the relevant magnetic interactions: 
\begin{equation}
\theta_{cw}=-\frac{S(S+1)}{3k_{B}}(2J_1+4J_2+2J_3)
\label{eq2}
\end{equation}
taking $J_{3}=49$~K, the triangle-based inter-chain couplings in \scto sum to $J_{inter}=J_1+2J_2=8$~K suggesting that they are antiferromagnetic and frustrated. As a result, \scto exhibits magnetic transitions at the temperatures $T_{N1}=5.5$~K, $T_{N2}=4.5$~K which are much lower than the Curie-Weiss temperature. They are revealed as peaks in the first derivative of the susceptibilities plotted in fig.~\ref{Fig2a}\textbf{c}. 

\begin{table}
\centering
\begin{tabular}{|c|c|c|c|} 
\hline
\bf Sample                      & \bf \begin{tabular}[c]{@{}l@{}}$\bm{\chi}_{vv}$ ($\times 10^{-5}$)\\(cm$^{3}$/mol)\end{tabular} & \bf g-factor & \bf \begin{tabular}[c]{@{}l@{}}$\bm{J_{chain}}$ ($J_{3}$)\\(K)\end{tabular} \\ 
\hline
Powder    & 3.85 $\pm$ 0.1 & 2.12 $\pm$ 0.005 & 49.1$\pm$0.02 \\ 
\hline
(100)     & 3.41 $\pm$ 0.1 & 2.19 $\pm$ 0.006 & 49.84$\pm$0.02 \\ 
\hline
(110)     &  1.59 $\pm$ 0.11 & 2.18 $\pm$ 0.006 & 50.09$\pm$0.02 \\
\hline
(111)     &  10.3 $\pm$ 0.13 & 2.15 $\pm$ 0.001 & 50.09$\pm$0.03 \\
\hline
\end{tabular}
\caption{The chain interaction strength and g-factor as derived by fitting the magnetic susceptibility above $T_{N1}$ ($T\ge$15~K, H$=0.05$~T) of the powder sample and single crystal sample aligned parallel to external field along the (100), (110) and (111) directions.}
  \label{tab2}
\end{table}

\begin{figure}
\includegraphics[width=1 \columnwidth]{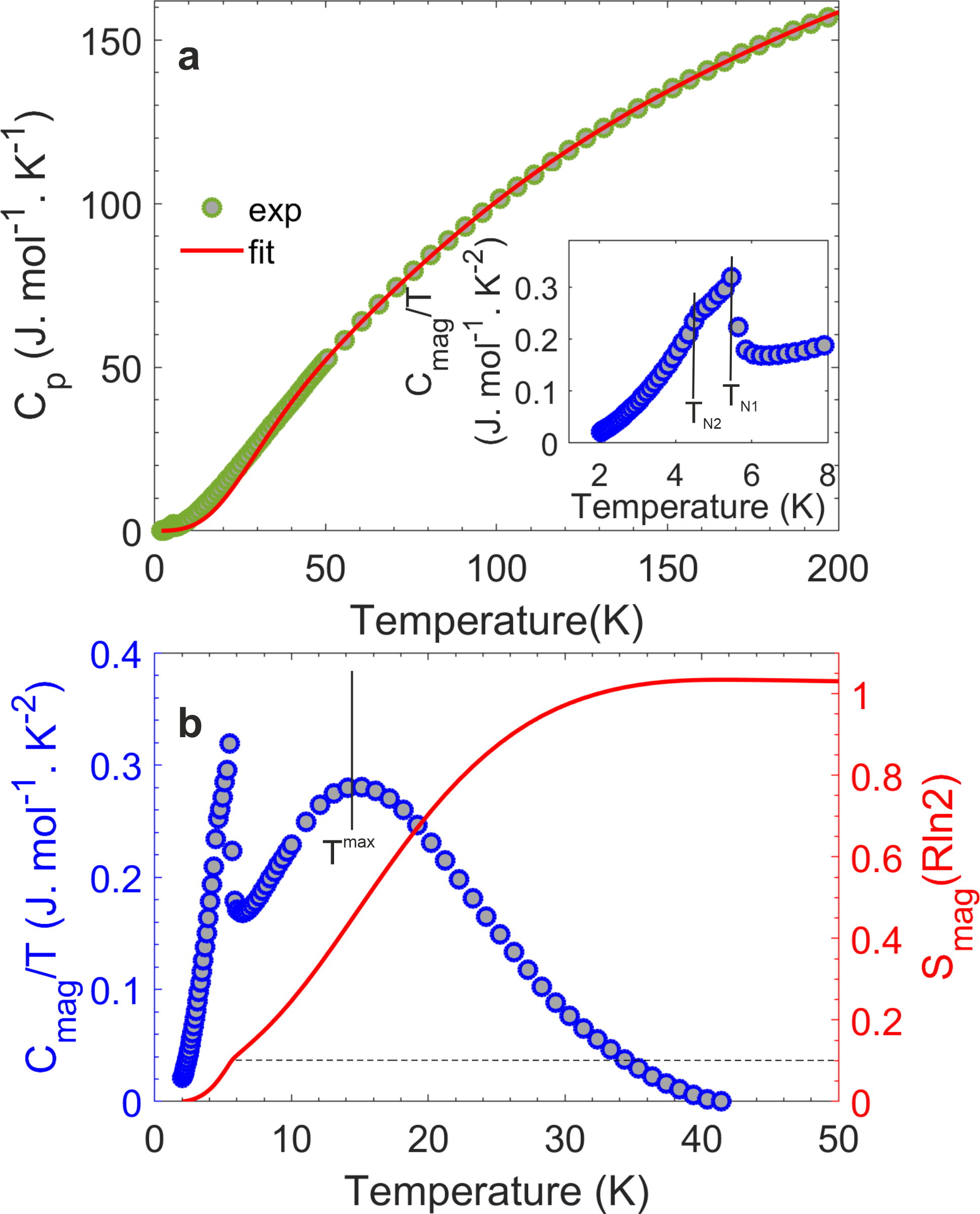}
\caption{\textbf{a}) Heat capacity of the crystalline sample. Red solid line is a fit to the Debye-Einstein model (eq.~\ref{eq3}) describing lattice heat capacity. Inset: $\lambda-$like anomalies at the two magnetic transitions at $T_{N1}=5.5$~K, $T_{N2}=4.5$~K. \textbf{b}) Left y-axis: the magnetic specific heat at low temperatures after subtracting the lattice contribution. Right y-axis: change in the magnetic entropy from the spin-1/2 value ($Rln2$) around the magnetic transition.}
\label{Fig2b}
\end{figure}

To confirm the presence of magnetic transitions, heat capacity of the single crystal has also been measured. As shown in the fig.~\ref{Fig2b}, the phonon contribution (C$_{phonon}$) of the high temperature heat capacity is very well described by a sum of one Debye integral and two Einstein terms given in eq.~\ref{eq3} (fit range 40~K$\ge$$T\ge$200~K) allowing the extraction of the dominant magnetic contribution at low temperatures.
\begin{equation}
\begin{split}
C_{phonon}(T)&=9R (n-C_{i})\left(\frac{T}{\theta_{D}}\right)^{3}\int_{0}^{\frac{\theta_{D}}{T}}\frac{x^{4}e^{x}}{(e^{x}-1)^{2}}dx\\
& + 3R\sum_{i=1,2}C_{i}\left(\frac{\theta_{E,i}}{T}\right)^{2}\frac{e^{\frac{\theta_{E,i}}{T}}}{(e^{\frac{\theta_{E,i}}{T}}-1)^{2}}
\end{split}
\label{eq3}
\end{equation}
Here, R$=8.3145$~J.~mol$^{-1}$.~K$^{-1}$ is the gas constant, $n$, $\theta_{D}$, $C_{i}$, $\theta_{E,i}$ are the no. of atoms per unit cell, Debye temperature, no. of Einstein modes and corresponding Einstein temperatures respectively.

\begin{figure*}
\includegraphics[width=2 \columnwidth]{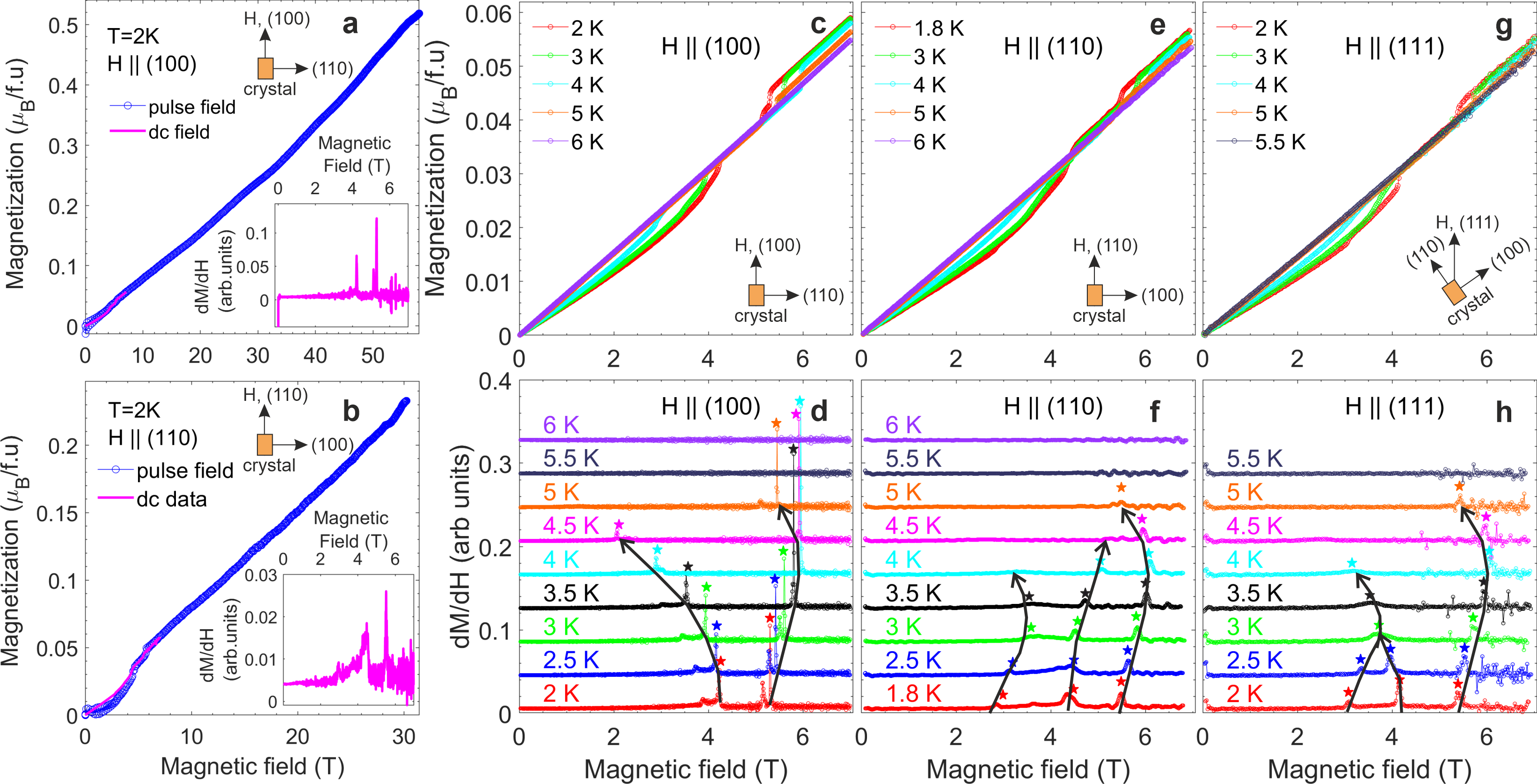}
\caption{\textbf{a-b}) Magnetization of \scto at 1.6~K measured in pulsed field and DC field at 2~K applied along the two crystalline directions (100) and (110) respectively. Insets: derivatives of magnetization measured in dc-field at 2~K. \textbf{c,e,g}) Magnetization curves measured at several temperatures in the dc-field for the three crystalline directions and, \textbf{d,f,h}) show the corresponding evolution of the derivatives of the magnetization indicating new field-induced transitions.}
  \label{Fig3b}
  \end{figure*}

The obtained magnetic quantity $C_{mag}/T$, where $C_{mag}=C_{p}-C_{phonon}$, shows two $\lambda$-like anomalies are observed at lower temperatures $T_{N1}=5.5$~K and $T_{N2}=4.5$~K (inset of fig.~\ref{Fig2b}a). These transitions are consistent with the previous reports in the polycrystalline samples. Above the magnetic transitions, $C_{mag}/T$ shows a broad peak at $\approx15.1$~K (left y-axis of fig.~\ref{Fig2b}b). This is a characteristic feature observed in Heisenberg spin-1/2 antiferromagnetic chains~\cite{,Klumper1998,Johnston2000} which relates to the chain interaction J$_{chain}$ as:
\begin{equation}
\frac{T^{max}_{C_{mag}/T}}{J_{chain}}\approx0.3072
\end{equation}
giving J$_{chain}=49.25$~K, in close agreement with the results from susceptibility. Although the magnitude of the magnetic contribution at higher temperatures varies with the fit range of the phonon contribution, we find that the magnetic entropy at lower temperatures ($\approx$ $T<10$~K) is unaffected by this artifact (right y-axis of the fig.~\ref{Fig2b}b). We find that only 10\% of the total magnetic entropy is released across the magnetic transitions ( $4.5$~K$<T<5.5$~K) . Therefore, the remaining 90\% of the entropy can be associated with the short range magnetic correlations corresponding to the one-dimensional nature of the Cu$^{2+}$ spins above the magnetic transition.

In order to explore the effects of magnetic field on \scto, magnetization measurements were performed at various temperatures. High field magnetization at $T=2$~K using a pulsed magnet, as well as lower field DC magnetization measurements along the (100) and (110) direction respectively are presented in fig.~\ref{Fig3b}\textbf{a}-\textbf{b}. The  pulsed field measurements were normalized by the DC magnetization and reveal that the Cu$^{2+}$ moment reaches 0.5~$\mu_{B}$ at 56~T. Considering a linear extrapolation, the saturation field can be expected at $\approx$ 110~T. 

At lower fields, two sets of anomalies are observed in the derivative of magnetization (in dc-field) along the (100) direction indicating possible field-induced magnetic transitions in the single crystal of \scto. As shown in the inset of fig.~\ref{Fig3b}\textbf{a}, these anomalies occur at $\approx$ 4.2~T and 5.5~T accompanied by shoulder peaks at 3.98~T and 5.13~T. Magnetization along crystalline (110) direction at 2~K (see inset of fig.~\ref{Fig3b}\textbf{b}) also reveals three anomalies at $\approx$ 3~T, 4.2~T and 5.5~T. These anomalies were followed as a function of temperature for the three directions of the single crystal (see fig.~\ref{Fig3b}\textbf{c, e, g})) as well as for the polycrystalline sample. The derivative of magnetization dM/dH in Fig.~\ref{Fig3b}\textbf{d} shows that the anomalies give rise to sharp and strong peaks when the field is applied along the (100) direction. With increasing temperature, the lower peak shifts to lower fields up to $T_{N2}=4.5$~K whereas the higher peak (5.5~T) shows a slight shift towards higher fields and disappears above 5~K. We observe that the shoulder peaks essentially move along with the main peaks. We believe this is due to a smaller crystallite within the sample with a misaligned (100) direction. 

\begin{figure*}
\includegraphics[width=1.95 \columnwidth]{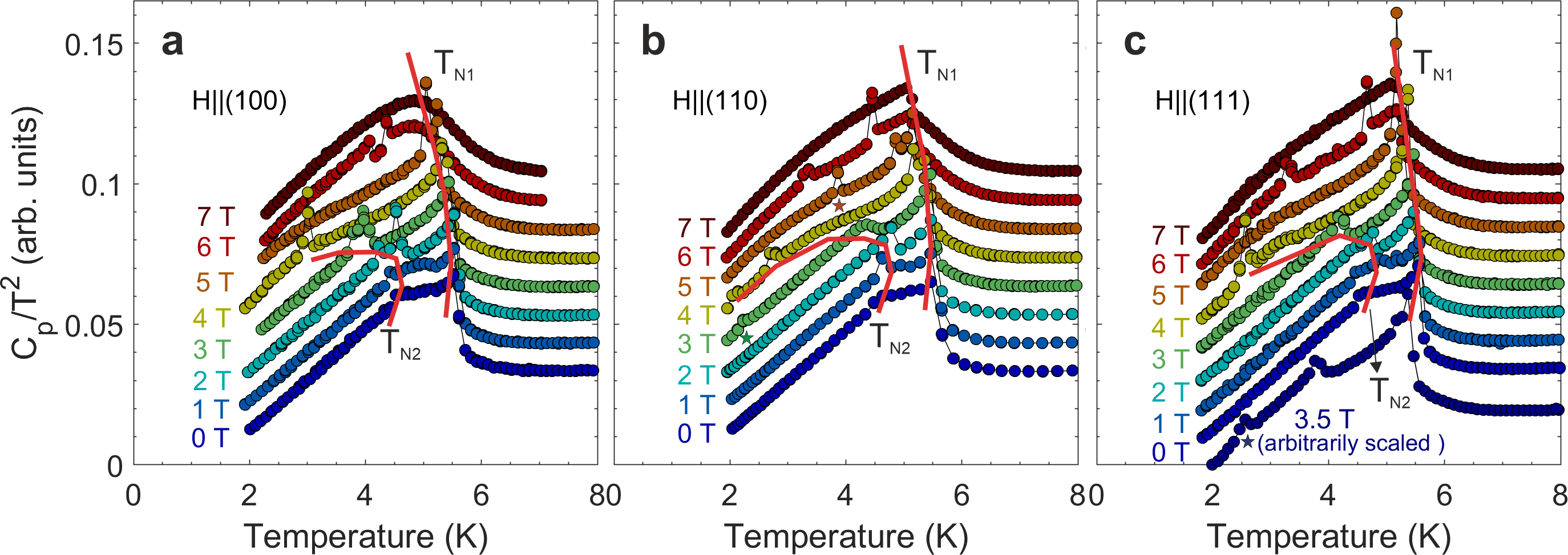}
\caption{Heat capacity $C_{p}/T^{2}$ of \scto as a function of temperature at several constant magnetic fields applied parallel to the crystalline \textbf{a}) (100), \textbf{b}) (110) and \textbf{c}) (111) directions. The additional stars in \textbf{b)-c)} indicate the additional anomalies compared to polycrystalline and (100) direction of the crystal.}
  \label{Fig3c}
  \end{figure*}

\begin{figure*}
\includegraphics[width=1.95 \columnwidth]{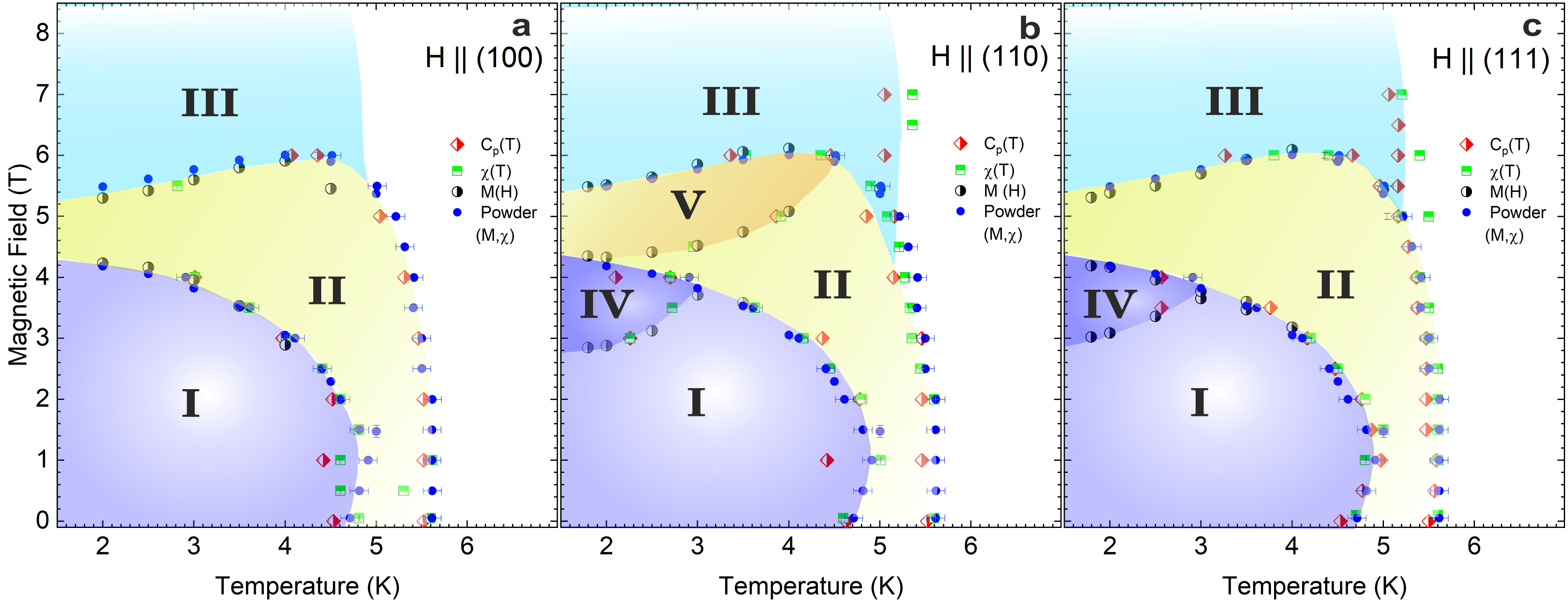}
\caption{H-T phase diagram of single crystal of \scto with external field applied \textbf{a}) along (100) direction, \textbf{b}) along (110) and \textbf{c}) along (111) directions.}
  \label{Fig4}
  \end{figure*}

Along the (110) direction, the peaks in the dM/dH are much weaker compared to the (100) direction, however, their position moves towards higher fields gradually up to $T_{N2}=4.5$~K where the highest field peak reaches a maximum of 6~T as shown in fig.~\ref{Fig3b}\textbf{f}. Only the highest field anomaly survives in the intermediate phase between $T_{N2}=4.5$~K and $T_{N1}=5.5$~K similar to the (100) direction. Finally, magnetization along the crystalline (111) direction (fig.~\ref{Fig3b}\textbf{g-h}) shows characteristics of behaviour along (110) as well as (100) direction. At base temperature $T=2$~K, the magnetization resembles mainly that of the (110) direction with anomalies in the dM/dH observed at $\approx$ 3.1~T, 4.1~T and 5.4~T. However, the two lower field anomalies merge at 3~K above which the peak shifts to lower fields and vanish above $T_{N2}=4.5$~K. On the other hand, the higher field anomaly stays between 5~T and 6~T similar to the other two directions.

These results are corroborated in the heat capacity measurements. The $\lambda$-like features corresponding to $T_{N1}$ and $T_{N2}$ in the specific heat also exhibit a significant field dependence in the three directions (see fig.~\ref{Fig3c}). We observe that the respective anomalies along (100) direction become sharper (indicated by solid red lines in fig.~\ref{Fig3c}\textbf{a}) in the external field. The $T_{N2}$ transition disappears above 4~T and a new transition anomaly is observed at 6~T. Above this field, a single, broad anomaly is seen at $T_{N1}$. While the behaviour of these transitions is similar along the (110) direction (fig.~\ref{Fig3c}\textbf{b}), two additional transition anomalies are observed at 2.1~K and 3.9~K in 3~T and 5~T field respectively (indicated by stars). These transitions are consistent with the anomalies observed in the magnetization of the crystal along (110) direction. The (111) direction of the crystal shows one additional peak at 2.6~K in 3.5~T field (blue star in fig.~\ref{Fig3c}\textbf{c}) while largely retaining the peaks corresponding to $T_{N1}$ and $T_{N2}$ from the (100) direction. However, the $T_{N1}$ transition remains sharp along (110) and (111) directions at fields H$\ge6$~T unlike along the (100) direction. Combining these observations, the phase diagram is then constructed for each of the crystal directions separately along with the polycrystalline sample.

Figure.~\ref{Fig4}\textbf{a} shows that phase diagram of the single crystalline \scto along (100) direction identifies three possible magnetic phases in the system. Here, phase-I refers to the magnetic ground state, phase-II is an intermediate phase and the phase-III, where heat capacity shows a broad $\lambda$, refers to ferromagnetic canting of the spins. These results are similar for the polycrystalline sample and in good agreement with the previously reported results~\cite{Koteswararao2015,Ahmed2015,Koteswararao2016}. Two additional phase-IV and phase-V are also observed when the field is applied along the (110) direction. Field along the (111) direction reveals phase-IV as well as the phases observed along the (100) direction. These additional phase transitions indicate a preferential orientation of the spins along the (110) direction which undergoes the most phase transitions whereas the presence (phase-IV along (111)) or absence (along (100)) of these additional phases could be attributed to the energy difference required to rotate the spins from (110) to (111) (35$^{\circ}$ rotation) or from (110) to (100) (55$^{\circ}$ rotation).

\subsection{Muon Spin Relaxation}
\begin{figure}
\includegraphics[width=0.99 \columnwidth]{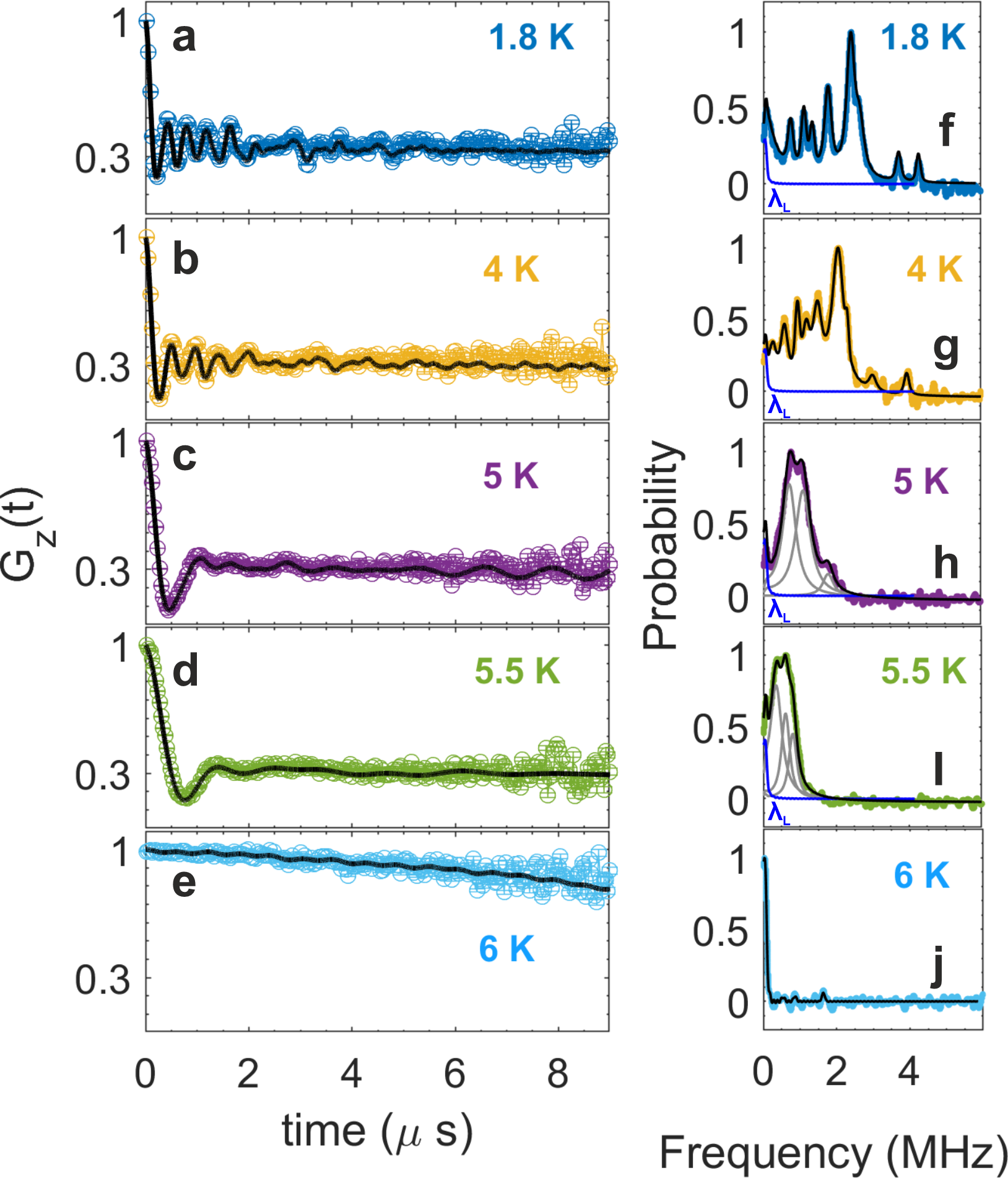}
\caption{\textbf{a}-\textbf{e}) Normalized $\mu$SR asymmetry spectra of powder \scto measured at GPS spectrometer in zero field as a function of temperature. The oscillations at the low temperature confirm the magnetic ordering and can be fitted (solid lines) with a 9-frequency component as described in the text. The corresponding Fourier transform of the $\mu$SR spectra (real part) are plotted in \textbf{f}-\textbf{j}). The multi-frequency model also describes the Fourier transform the oscillations very well as shown by the black solid lines. The blue solid lines indicate non-oscillating dynamic part decaying with $\lambda_{L}$ rate. The grey solid lines in \textbf{h}-\textbf{i} represent the three Gaussian terms in the intermediate phase.}
  \label{Fig5}
  \end{figure}

To obtain more insight into the nature of the magnetic order below the two transitions $T_{N1}$ and $T_{N2} $ in \scto we further probe the material with muon spin relaxation ($\mu^{+}$SR) experiments in zero magnetic field between 2~K and 10~K. Figure.~\ref{Fig5}\textbf{a-e} show the $\mu^{+}$SR spectra of \scto as a function of decay time at several temperatures in the ordered state ($T<T_{N1}=5.5$~K) and in the paramagnetic state $T=6$~K. At base temperature, the spectrum clearly reveals the oscillatory behavior of the asymmetry resulting from the Larmor precession of the muon spin around the local internal field set by the magnetic ordering in the system. Furthermore, the remnant relaxation at long time-scales saturates at $\frac{1}{3}$ of the initial value of the asymmetry. These observations are typical indications of static magnetic order in the system. 

\begin{figure}
\includegraphics[width=1 \columnwidth]{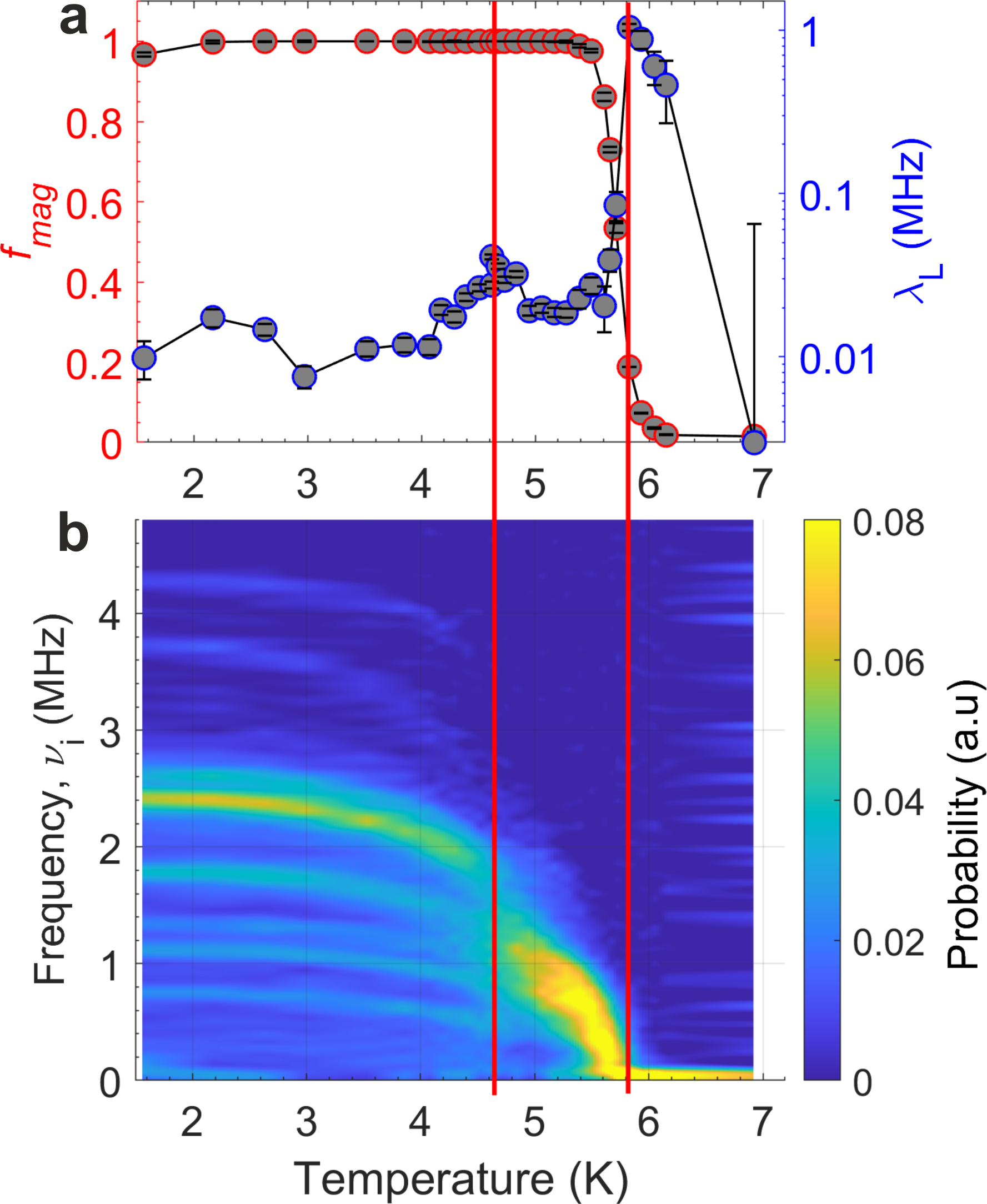}
\caption{\textbf{a}) Left yaxis: magnetic fraction $f_{mag}$ as described in the eq.~\ref{eq5}. Right y-axis: Longitudinal relaxation $\lambda_{L}$ of the $\mu$SR spectra and, \textbf{b}) map of the Larmor precession frequencies, proportional to the order parameter, below the magnetic transitions in polycrystalline \scto.}
\label{Fig6}
\end{figure}

The Fourier transform (FFT) of the oscillating spectra reveals nine frequency components at base temperature as shown in fig.~\ref{Fig5}\textbf{f} and their distribution varies as the temperature increases towards $T_{N2}$ (fig.~\ref{Fig5}\textbf{f}-\textbf{g}). Therefore, all the spectra below $T_{N2}$ are fitted by considering a superposition of nine Gaussian-distributed internal magnetic fields to describe the precessing part of the spectrum as described in the following model:

\begin{equation}
\begin{split}
G_{z}(t)&=f_{mag} \Biggl[\frac{2}{3}\sum_{i=0}^{9} A_{T,i}Cos(2\pi\nu_{i}t)e^{-\lambda_{T,i}t} + \frac{1}{3}e^{-\lambda_{L}t}\Biggr] \\
&+(1-f_{mag}) G_{KT} e^{-\lambda_{bkg}t}
\end{split}
\label{eq5}
\end{equation}

where $G_{KT}$ is the Gaussian Kubo-Tayabe function that describes the asymmetry due to nuclear moments in the paramagnetic state and $f_{mag}$ is the magnetic contribution due to electronic spin ordering in the system. The magnetic part is further separated into $\frac{2}{3}$ Cosine-oscillating term consisting of nine frequency contributions ($\nu_i$) with weight fractions $A_{T,i}$, and $\frac{1}{3}$ non-oscillating relaxing term at long time-scales. The former term describes a homogeneous Gaussian distribution of internal fields and the latter term implies the relaxation ($\lambda_{L}$) of those muons whose spin is longitudinal to the internal field at the time of decay and hence is indicative of the spin dynamics in the system. Upon approaching $T_{N1}$ from high temperatures the magnetic fraction $f_{mag}$ reaches a value of unity (left y-axis of fig.~\ref{Fig6}\textbf{a}) confirming that all of the Cu$^{2+}$ in \scto undergo magnetic transition and eliminating the possibility of phase separation. Consequently, $\lambda_{L}$ peaks up at $T_{N1}=5.8$~K and $T_{N2}=4.6$~K and as shown in the right y-axis of fig.~\ref{Fig6}\textbf{a} reflecting the critical dynamics at the magnetic transitions in \scto. These transition temperatures are in close agreement with the values observed in the magnetic heat capacity and susceptibility measurements.  

The field distribution below $T_{N2}$ is clearly separated into nine components (as explained above) with the strongest frequency at base temperature occurring at $\nu$=2.4~MHz. This refers to an internal field of 0.18~kOe with a small field distribution (gaussian width) of $\Delta\nu=0.729$~MHz$=5$~mOe. Above $T_{N2}$, the nine frequency components collapse into a broad peak (fig.~\ref{Fig6}\textbf{b}). To further understand the distribution of the field in this region two spectra, namely 4.8~K and 5~K, have been fitted by considering one, two and three Gaussian terms respectively with 3-Gaussian (fig.~\ref{Fig5}\textbf{h,i}) resulting in a best fit. This model also sufficiently describes all the temperatures between $T_{N1}<T<T_{N2}$ ($\chi^{2}\approx1$). For consistency, the broad field distribution in this range has also been analyzed using a zeroth order Bessel function that points to an incommensurate spin density wave model~\cite{Savici2002}. The resulting $\chi^{2}$ was found to be 2.6 clearly indicating that the model is not applicable in \scto. With increasing temperature the broad Gaussian gradually moves to smaller frequencies and completely vanishes above the highest transition at $T_{N1}=5.8$~K. 

We may attribute the origin of these frequencies to a composite of the muon sites around three inequivalent Oxygen sites (Tab.~\ref{tab0}) (with three Cu-O bond lengths: 1.939~\AA, 1.943~\AA{} and 3.086~\AA{}) and local spin directions of the 12 Cu moments with respect to the incoming $\mu^{+}-$ spin. However, a confirmation of the same requires a detailed calculation of muon sites based on the Coulomb potential. Nevertheless, the ZF-$\mu$SR data clearly reveal two different magnetic phases with distinguishing internal field distributions in zero-field.

\subsection{Magnetic structure}\label{sec:magstruct}

\begin{figure}
\includegraphics[width=1 \columnwidth]{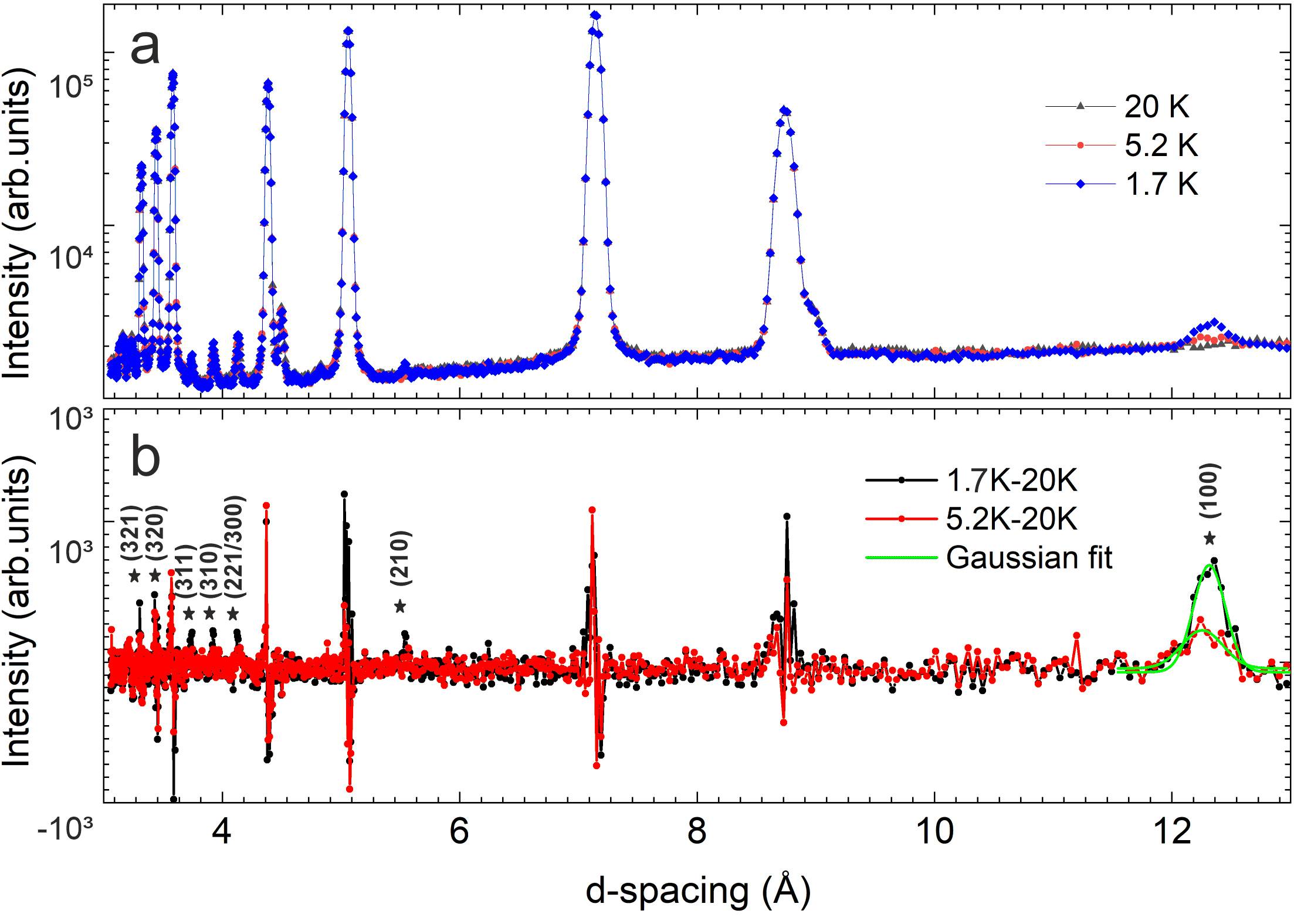} 
\caption{\textbf{a}) Powder neutron diffraction patterns measured at the DMC diffractometer below the magnetic transitions at 1.7~K, 5.2~K and above at 20~K. \textbf{b}) The difference patterns with respect to 20~K reveal several magnetic peaks. The solid green lines refer to Gaussian fit of the (1,0,0) peak at 12.43~\AA{} for the two subtracted patterns.}
  \label{Fig7}
  \end{figure}

To investigate the magnetic structure of \scto in the ground state, i.e., below $T_{N2}=4.5$~K, several powder diffraction patterns are obtained between temperatures 1.7~K and 7~K. Representative low temperature diffraction patterns of \scto obtained on the DMC diffractometer are plotted in fig.~\ref{Fig7}\textbf{a} for a polycrystalline sample at the base temperature 1.7~K, in the intermediate magnetic phase at 5.2~K and in the paramagnetic state at 20~K. These patterns reveal that the nuclear structure of the \scto remains unchanged even below the magnetic transition. Additionally, a new Bragg peak is observed at $d=12.3304$~\AA{} corresponding to the (1,0,0) reflection below the magnetic transition at 1.7~K. The patterns subtracted from data at high temperature (see fig.~\ref{Fig7}\textbf{b}) clearly shows that the (100) peak survives even at 5.2~K. Furthermore, Gaussian fit of the peak (solid green line in fig.~\ref{Fig7}\textbf{b}) reveals that its position and FWHM remain unchanged within the error bars at the two temperatures (0.41 $\pm$ 0.08~\AA{} and 0.32 $\pm$ 0.03~\AA{} respectively for 5.2~K and 1.7~K). The subtracted patterns also reveal additional magnetic intensities clearly visible on the weak nuclear peaks (2,1,0), (3,0,0)+(2,2,1), (3,1,0) and (3,1,1) at the d-spacing of 5.6~\AA, 4.2~\AA, 4~\AA{} and 3.8~\AA{} respectively. However, the contribution of magnetic intensity on the strong nuclear peaks is ambiguous. Although the structural peaks at (2h+1,0,0) are allowed for the primitive type of unit cell, the four-fold screw symmetry of space group {\it P4$_{1}$32} forbids these peaks while allowing only those with h$=4$n. Therefore, the magnetic propagation vector can be identified as $q_m=(0,0,0)$.

\begin{figure}
\includegraphics[width=1 \columnwidth]{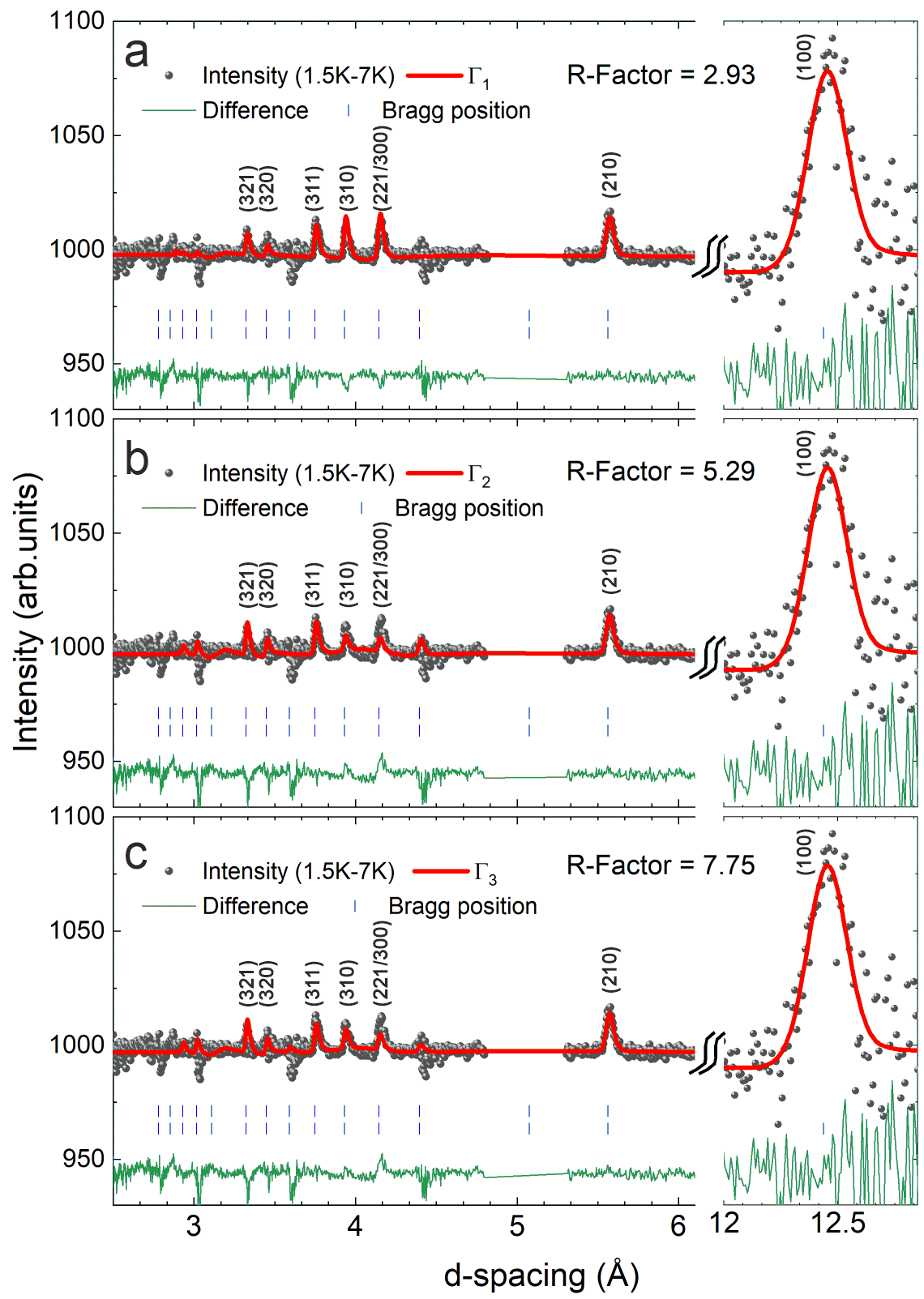}
\caption{\textbf{a-c}) Rietveld refinement of the magnetic intensities measured at the WISH diffractometer at 1.6~K (obtained by subtracting the intensity at 7~K) using three different irreducible representations of the magnetic structure for \scto.}
  \label{Fig8}
  \end{figure}

Representation analysis for the propagation vector (0,0,0) reveals that the reducible magnetic representations ($\Gamma_{mag}$) associated with the 12d Wyckoff position of Cu decomposes into direct sum of five irreducible representations (IRs) denoted as $\Gamma_i$ ($i=1-5$). We use superscript to indicate dimensionality of the IRs:

\begin{equation}
\Gamma_{mag}=1\Gamma_{1}^{1}+2\Gamma_{2}^{1}+3\Gamma_{3}^{2}+4\Gamma_{4}^{3}+5\Gamma_{5}^{3}
\label{eq4}
\end{equation}

Following the standard approach, the solution of the magnetic structure was searched assuming a single IR (irreducible magnetic order parameter). For the three-dimensional IRs $\Gamma_4$ and $\Gamma_5$, only high-symmetry combinations of the basis functions corresponding to maximal isotropy subgroups~\cite{Campbell2006}, were tested. The low-symmetry magnetic structures require a strongly first order phase transition and are unlikely from the thermodynamic point of view. The systematic absence of the (2h,0,0) magnetic reflections is inconsistent with the $\Gamma_4$ and $\Gamma_5$ IRs, while discrimination between $\Gamma_1$, $\Gamma_2$ and $\Gamma_3$ were more challenging. As the changes on the strong nuclear peaks such as (1,1,0), (1,1,1) and (2,1,1) are not clear, these regions are excluded from the analysis while refining the magnetic structure. For this we used high intensity datasets collected on the WISH time-of-flight diffractometer. The magnetic intensity was obtained by subtracting the 7K data from the 1.5~K dataset. 

\begin{figure}
\includegraphics[width=0.9 \columnwidth]{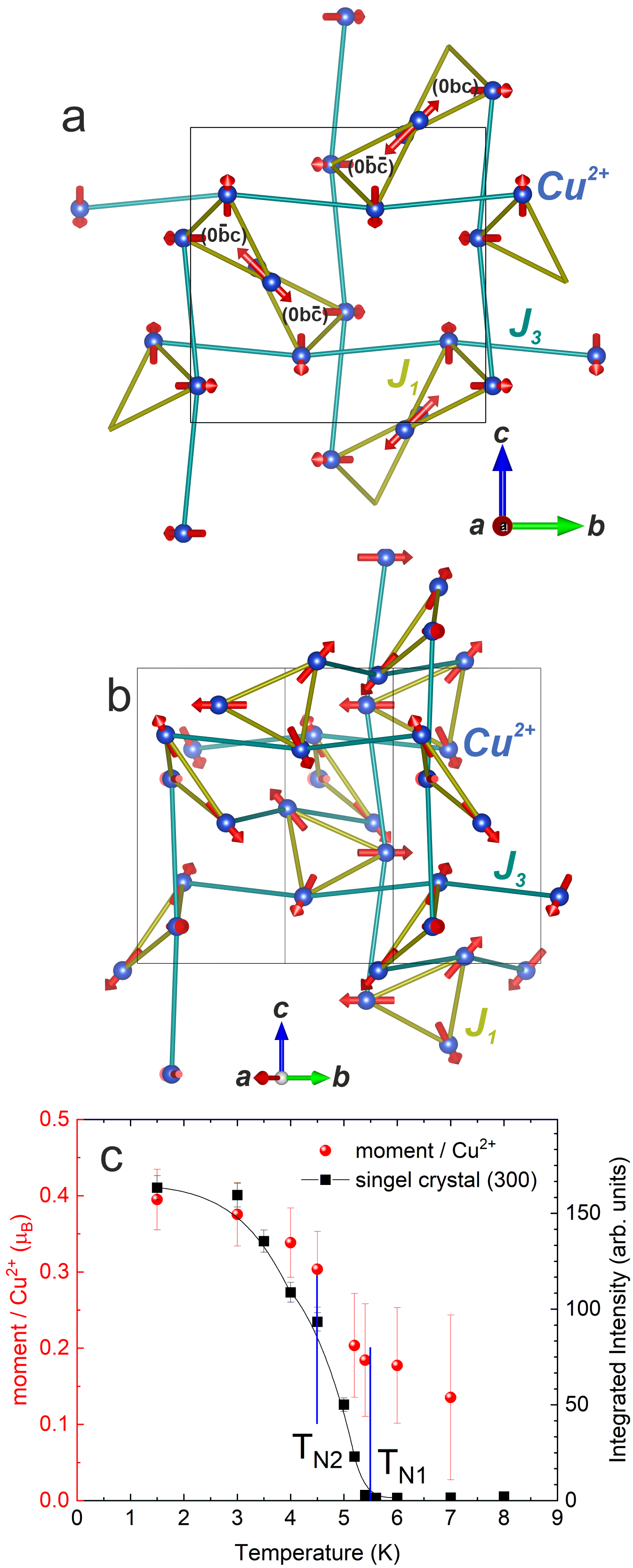}
\caption{\textbf{a} Magnetic structure of \scto described by $\Gamma_{1}$ representation at the base temperature 1.7~K showing the two chains propagating along each of the cubic axes within a single unit cell. Spins in a chain are perpendicular to those in the neighboring parallel chain in the same direction. \textbf{b} shows the inter-chain coupling promoted by first nearest neighbour interaction $J_{1}$, \textbf{c} the temperature dependence of the ordered moment refined on the polycrystalline sample by considering the magnetic structure from $\Gamma_{1}$ as well as the integrated intensity of the magnetic peak (3,0,0) of the single crystalline \scto below 7~K.}
  \label{Fig9}
  \end{figure}

Figure.~\ref{Fig8}\textbf{a-c} show individual refinements of the magnetic peaks for IRs $\Gamma_1$, $\Gamma_2$ and $\Gamma_3$ respectively. All the three representations reproduce the strongest magnetic peak (100) (at d=12.438\AA) very well with the differences in fit quality appearing only at high-Q peaks such as (2,2,1)+(3,0,0) ($d=3.933$~\AA) and (3,1,0) ($d=4.1461$~\AA) resulting in a best magnetic Bragg-factor (2.93) from the first IR, $\Gamma_{1}$. The corresponding magnetic structure implies the cubic magnetic symmetry P4$_1$32 ($\#$213.63) with the basis and origin defined with respect to the paramagnetic space group as: (1,0,0), (0,1,0), (0,0,1) and (-1/4,-1/4,-1/4), respectively. In this magnetic structure, each of the Cu-spins is aligned along a local (1,1,0) direction. Here, the third nearest neighbours of Cu$^{2+}$ forms antiferromagnetic spin-$\frac{1}{2}$ chains running along the three mutually perpendicular crystallographic {\it a-}, {\it b-} and {\it c-} axes. Furthermore, we observe two parallel chains per cubic direction, as shown in fig.~\ref{Fig9}\textbf{a} for chains along {\it a-} axis, whose spins take on two perpendicular spin directions in the {\it b-c} plane, (0,1,1) and (0,1,-1). This results in a total of 6 spin directions in the ordered state of \scto so that the frustrated first nearest-neighbour interaction $J_{1}$ forms co-planar 120$^{\circ}$ triangles as shown fig.~\ref{Fig9}\textbf{b}. Although these triangles are isolated from each other, spins on the vertices of the every triangle participates in coupling the three perpendicular spin-chains leading to three dimensional magnetic order in the system. It is clear that the $J_{1}$ rather than the hyperkagome interactions $J_{2}$, are responsible for the inter-chain coupling.

We observe that the magnetic propagation vector remains unchanged even in the intermediate phase within the instrumental resolution. Therefore, the pattern in this temperature range is also refined by the same magnetic structure resulting from $\Gamma_{1}$. Figure.~\ref{Fig9}\textbf{c} shows the evolution of Cu$^{2+}$ moments as a function of temperature which reaches a maximum ordered moment of $\sim$0.4$~\mu_B$ at 1.6~K. 
The total ordered moment calculated by Schulz {\it et al.,}~\cite{Schulz1996} for Heisenberg spin-1/2 chain with interchain coupling J$_{inter}$ using mean-field-theory is given as:

\begin{equation}
m_0=1.0197\sqrt{\left(\frac{J_{inter}}{J_{chain}}\right)}
\end{equation}

which yields a value of $m_0\approx0.41~\mu_B$ for \scto considering $J_{inter}=8$~K and $J_{chain}=50$~K. While this value is consistent with the experimental moment at the base temperature, it also confirms the presence of weak antiferromagnetic inter-chain coupling responsible for the loss of 60\% of full moment expected for fully ordered Cu$^{2+}$ spin as would be found in a 3D ferromagnet. As the error bars of the moment obtained from powder diffraction are high, we have also followed the intensity of the magnetic Bragg peak (300) in the single crystal of \scto (right y-axis of fig.~\ref{Fig9}\textbf{c}) which clearly indicates a non-zero intensity below the first magnetic transition $T_{N1}=5.5$~K. However, no significant changes are observed at the lower transition $T_{N2}=4.5$~K.     

\section*{Discussion}

The magnetic, thermodynamic properties and $\mu^{+}$SR measurements clearly identify two magnetic phases in \scto in zero field at $T_{N1}\approx$5.5~K and $T_{N2}\approx$4.5~K. The low temperature phase (Phase-I in fig.~\ref{Fig4}) below $T_{N2}$ is described by a co-planar 120$^{\circ}$ structure of the Cu spins coupling three mutually perpendicular AF chains so that each of the spins points along a local (110) direction as explained in the sec.~\ref{sec:magstruct}. The intermediate phase (phase-II in fig.~\ref{Fig4}) between $T_{N1}$ and $T_{N2}$ is associated with broad local field distribution around the muon site. However, we note that there is no indication for an incommensurate spin structure as the field distribution is always Gaussian-like pointing to a homogenous local internal field instead of continuous fields centered around 0~T expected for a helical/chiral spin structure or spin density wave type of modulation~\cite{Savici2002,Yaouanc2017}.

The preferential local (110) direction of the spin structure in the ground state is also apparent in the H-T phase diagram. When the field is applied along (110) direction i.e., parallel to one of the local ordered spin directions, five different phases can be identified. Whereas field along (111) and (100) result in four and three phases respectively as shown in fig.~\ref{Fig4}. While heat capacity data reveals sharp peaks at the phase boundaries in all the three directions (see see fig.~\ref{Fig3c}), the changes in magnetization are sharpest along (100) direction (see fig.~\ref{Fig3b}) and weakest along the (110) direction suggesting that the latter is also a preferred magnetization direction. Additionally, the presence of phase-IV along (111) also reveals its component along the preferred (110) direction. However, the boundary of the paramagnetic phase (above $T_{N1}$) to phase-III in all the three directions is weak compared to that of paramagnetic to phase-I revealing that phase-III consists of weak ferromagnetic behaviour due to canting of the spins along applied field.

The small ordered moment in the ground state (only 40\% of each spin orders in zero field) indicates that the spins are either highly frustrated or highly one dimensional. If the former, strong frustration would imply a strong hyperkagome interaction $J_2$ which would be incompatible with the antiferromagnetic alignment in the chains and an incommensurate magnetic order might be expected in the ground state. However, the $\mu$SR and diffraction experiments rule out this possibility. Furthermore, we find that only 10\% of the magnetic entropy is released at the magnetic transition while the other 90\% is recovered below $\approx$ 40~K where one-dimensional magnetism is relevant, revealing that the $J_2$ is weak and possibly its net effect is cancelled. Whereas in the latter case, the chain interaction $J_3$ is strong and dominates the magnetic structure giving rise to the antiferromagnetic chain, while the weaker triangle interaction $J_1$ which is compatible with this order, couples mutually perpendicular chains together into a 120$^{\circ}$ spin arrangement. 

This observation is clearly in contrast to the strong frustration observed in \pcto due to the hyper-hyperkagome interactions (where the $J_1$ and $J_2$ interactions are dominant, antiferromagnetic and of equal strength.)~\cite{Chillal2020} despite the structural similarity. However, some differences between these two compounds still remain in the form of bond angles responsible for the super-exchange pathways as proposed by Koteswararao \textit{et al.}~\cite{Koteswararao2015}. For instance the ratio of bond angles responsible for $J_2$ (Sr: 92.5$^{\circ}$, Pb: 97$^{\circ}$) and $J_3$ (Sr: 162.2$^{\circ}$, Pb: 156$^{\circ}$), J$_2-$angle/J$_3-$angle, is $\approx$ 9\% higher in \pcto compared to \scto. In addition, the extra lone-pair in \pcto might play a key role in the weaker chain interaction due to the hybridization of the Pb-O bonds, involved in the $J_3$ superexchange path (O-Pb2-O), that may have extra strain effects as in ferroelectric perovskite systems~\cite{Cohen1992}. Confirmation of this needs a detailed investigation into the electronic band structure of both the systems, which is beyond the scope of this work. 

Koteswararao et al.~\cite{Koteswararao2016} find magnetoelectric effects in the form of electric polarization at magnetic transitions in \scto in an applied magnetic field manifesting a strong coupling between magnetism and lattice. The field-induced polarization also resulted in a similar phase diagram as that of the magnetic phase transitions observed in polycrystalline and crystalline (100) directions as a function of field. It would therefore not be surprising if antiferromagnetic order also influenced the structure so that structural changes occur at the transitions to the long-range magnetic order. These changes are likely to be much smaller in zero field such as symmetry allowed displacements which retain the nuclear space group. Hence, no visible changes were observed on the nuclear peaks in the powder diffraction patterns. However, heat capacity results in field (see fig.~\ref{Fig3c}) reveal a sharper $\lambda$-anomaly above 3~T at $T_{N1}$, consistent with the field induced electric polarization. Therefore, investigation of magnetic structure of \scto in an external field would give insight into the origin of the spin-lattice coupling.  

\section*{Summary}
In summary, we have studied magnetic properties of \scto in polycrystalline and single crystal samples and investigated the magnetic structure. The field-dependent phase diagram in single crystals reveals additional magnetic phases for the (110), (111) directions whereas the (100) direction replicates the phase diagram of the polycrystalline sample. We propose a magnetic structure of \scto where, $J_{1}$ acts as an inter-chain coupling to the AF chains formed by $J_{3}$ leading to three dimensional magnetic ordering in the system below $T_{N1}$. 

Note: As this paper was being finalized we became aware of a similar investigation of \scto on arXiv~\cite{saeaun2020}. While there are some differences in the techniques employed, the results of that paper are in broad agreement with this paper.

\section*{Acknowledgements}
S.C thanks M. Reehuis for discussion on the magnetic symmetry analysis, J. Schr\"oter, K. Siemensmeyer and R. Feyerherm for supporting the thermodynamic and magnetometry measurements. B.L acknowledges the support of the Deutsche Forschungsgemeinschaft (DFG) through the project B06 of the SFB-1143 (ID:247310070). We acknowledge the support of the HLD at HZDR, member of the European Magnetic Field Laboratory (EMFL). 
\bibliographystyle{apsrev4-2}

\end{document}